\documentclass[twocolumn,preprintnumbers,amsmath,amssymb,prb]{revtex4}

\usepackage{graphicx}
\usepackage{dcolumn}
\usepackage{bm}
\usepackage{color}

\begin{document}

\preprint{}

\title{Magnetoelastic effects in the metallic frustrated antiferromagnet CrB$_2$}

\author{Tadataka Watanabe$^1$}
\email{watanabe.tadataka@nihon-u.ac.jp}
\author{Mai Watanabe$^1$}
\author{Sakurako Suganuma$^1$}
\author{Andreas Bauer$^{2,3}$}
\author{Christian Pfleiderer$^{2,3,4,5}$}
\affiliation{$^1$Department of Physics, College of Science and Technology, Nihon University, Chiyoda, Tokyo 101-8308, Japan}
\affiliation{$^2$Physics Department, TUM School of Natural Sciences, Technical University of Munich, D-85748 Garching, Germany}
\affiliation{$^3$TUM Center for Quantum Engineering (ZQE), Technical University of Munich, D-85748 Garching, Germany}
\affiliation{$^4$Munich Center for Quantum Science and Technology (MCQST), D-80799 Munich, Germany}
\affiliation{$^5$Heinz Maier-Leibnitz Zentrum (MLZ), Technical University of Munich, Lichtenbergstr. 1, D-85747 Garching, Germany}
\date{\today}

\begin{abstract}
Hexagonal chromium diboride CrB$_2$ is a metallic frustrated antiferromagnet with a N\'{e}el temperature $T_N \sim$ 88 K. In CrB$_2$, Cr 3$d$ electrons not only give rise to localized magnetic moments but also contribute to metallic conduction. We perform ultrasound velocity measurements on a single crystal of hexagonal CrB$_2$ to determine its elastic properties. The temperature dependence of the $ab$-plane shear elastic modulus exhibits Curie-type softening upon cooling from $\sim$120 K down to $T_N$. This behavior is interpreted as a precursor to a symmetry-lowering lattice distortion at $T_N$, indicating that magnetic frustration is relieved via transverse magnetoelastic coupling. In addition, the $a$-axis and $c$-axis compressive elastic moduli show unusual softness and their suppression upon cooling, which are naturally explained by Fermi-surface nesting and its suppression upon cooling. The present results suggest that, in CrB$_2$, longitudinal magnetoelastic coupling suppresses Fermi-surface nesting and enhances frustrated exchange interactions, while transverse magnetoelastic coupling plays a key role in relieving the frustration.
\end{abstract}

\maketitle

\section{Introduction}

Magnetically frustrated systems provide a rich platform for exploring unconventional correlated phenomena arising from ground-state degeneracies [\cite{Lacroix1}]. For frustrated magnets, the majority of studies have been devoted to insulating materials, where the magnetism is dominated by nearest-neighbor (and next-nearest-neighbor) exchange interactions. There are comparatively fewer studies on metallic frustrated magnets [\cite{Lacroix2,Vojta}]. In metallic magnets, the coupling of conduction electrons with local moments can generate further-neighbor and higher-order exchange interactions and can cause unique frustration phenomena.

Chromium diboride CrB$_2$ is an antiferromagnetic (AF) metal with an ordering temperature $T_N \sim$ 88 K [\cite{Bauer}]. Its Curie--Weiss behavior, characterized by an AF Weiss temperature $\theta_W \sim -750$ K and a large effective moment $\mu_{eff}\sim$ 2.0 $\mu_B$ f.u.$^{-1}$, indicates the presence of frustration among Cr spins, as reflected in the large ratio $|\theta_W/T_N|\sim$ 8.5 [\cite{Bauer}]. Furthermore, resistivity and specific heat measurements reveal that CrB$_2$ is a weak itinerant antiferromagnet with strong AF spin fluctuations characterized by a spin fluctuation temperature $T_{SF} \sim$ 257 K [\cite{Bauer}]. CrB$_2$ crystallizes with an AlB$_2$-type hexagonal structure consisting of a Cr triangular lattice stacked along the $c$-axis [Fig. 1] and an alternately stacked B honeycomb lattice. For CrB$_2$, earlier neutron scattering experiments on the AF phase identified an incommensurate cycloidal magnetic order with a propagation vector {\bf q} = (0.285,0.285,0) and a reduced ordered magnetic moment of 0.5 $\mu_B$ f.u.$^{-1}$, which is characteristic of itinerant magnetism [\cite{Funahashi,Kaya}]. In addition, a recent inelastic neutron scattering study revealed that the spin-wave spectra observed in the AF phase can be described by a spin Hamiltonian incorporating long-range Ruderman--Kittel--Kasuya--Yoshida (RKKY) interactions including further-neighbor couplings within the $ab$-plane [\cite{Park}]. Band structure calculations for CrB$_2$ suggest that the most significant contributions come from Cr 3$d$ orbitals near the Fermi surface, along with B 2$p$ orbitals [\cite{Liu,Wang,Vajeeston,Brasse,Biswas,YWang}]. Consequently, it is believed that in CrB$_2$, 3$d$ electrons not only create localized magnetic moments but also contribute to metallic conduction.

For frustrated CrB$_2$, thermal expansion measurements reveal the presence of magnetoelastic coupling [\cite{Nishihara}], indicating that a symmetry-lowering lattice distortion driven by magnetoelastic coupling can relieve the frustration, resulting in the formation of cycloidal AF order with {\bf q} = (0.285,0.285,0) [\cite{Funahashi,Kaya}]. In a frustrated system with magnetoelastic coupling, the ground-state degeneracy can be lifted by a symmetry-lowering lattice distortion [\cite{Yamashita,Tchernyshyov}]. This effect is called the spin Jahn--Teller (spin-JT) effect because its mechanism is analogous to the Jahn--Teller effect in orbital-degenerate systems where spontaneous lattice distortion reduces crystal symmetry to lift the orbital degeneracy [\cite{Gehring}].

The spin-JT effect was initially proposed for cubic pyrochlore antiferromagnets in which spins form a lattice of corner-sharing tetrahedra [\cite{Yamashita,Tchernyshyov}]. For real pyrochlore antiferromagnets, the spin-JT mechanism has been applied to explain the magnetostructural transitions in insulating cubic spinels $A$Cr$_2$O$_4$ ($A$ = Mg, Zn, Cd), where a cubic-to-tetragonal lattice distortion relieves magnetic frustration [\cite{Lee,Ortega-San-Martin,Chung,Watanabe3}]. Recently, insulating orthorhombic pseudobrookite CoTi$_2$O$_5$, which has much lower crystal symmetry than the prototypical spin-JT system $A$Cr$_2$O$_4$, has been experimentally identified as a spin-JT antiferromagnet [\cite{Kirschner,Watanabe10,Guratinder}]. In CoTi$_2$O$_5$, the symmetry-lowering lattice distortion below the AF ordering temperature $T_N$ is too small to be resolved experimentally [\cite{Kirschner}]. CoTi$_2$O$_5$ has been confirmed to be a spin-JT antiferromagnet through the observation of anomalies in its elastic moduli [\cite{Watanabe10}] and acoustic phonons [\cite{Guratinder}]. For CrB$_2$, although the symmetry-lowering lattice distortion in the AF phase has not yet been experimentally resolved, there remains the possibility of the emergence of a spin-JT effect.

\begin{figure}[t]
\begin{center}
\includegraphics[scale=0.55]{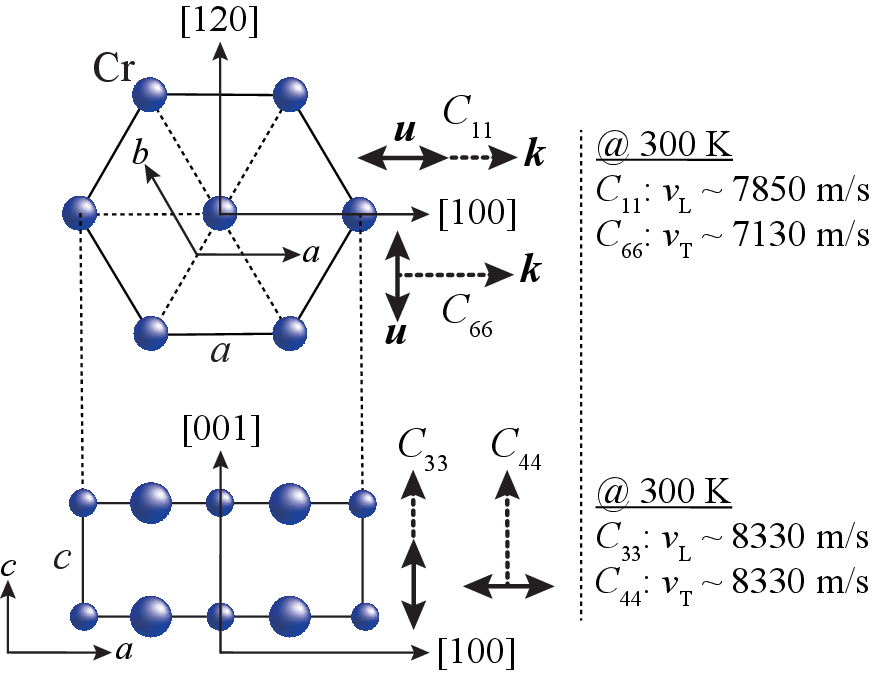}
\caption{\label{fig:fig1} Cr sites of CrB$_2$ in the hexagonal crystal unit cell, with the propagation vector {\bf k} and polarization vector {\bf u} of the sound waves for the compressive elastic moduli $C_{11}$ and $C_{33}$, and the shear elastic moduli $C_{44}$ and $C_{66}$. The sound velocities of the longitudinal ($v_L$) and transverse ($v_T$) waves corresponding to these elastic moduli of CrB$_2$ measured at a temperature of 300 K are listed on the right.}
\end{center}
\end{figure}

For metallic CrB$_2$, earlier band structure calculations suggested the presence of Fermi-surface (FS) nesting along the $c$-axis and predicted the emergence of spin-density-wave (SDW) order [\cite{Liu}]. Although this SDW order is different from the experimentally identified cycloidal AF order with {\bf q} = (0.285,0.285,0) [\cite{Funahashi}], the possibility of FS nesting and phonon softening arising from the Kohn anomaly still remains [\cite{Kohn}]. Recent band structure calculations have suggested the presence of three-dimensional FS nesting and predicted the emergence of pressure-induced hybrid $s$-wave superconductivity resulting from the competition between spin fluctuations and electron--phonon coupling [\cite{Biswas}].

In this paper, we present ultrasound velocity measurements of hexagonal CrB$_2$, from which we determine its elastic moduli. The sound velocity, which is closely related to elastic modulus, is a useful probe that enables symmetry-resolved thermodynamic information to be extracted from a crystal [\cite{Luthi}]. In magnets, magnetoelastic coupling can modify the dispersion of sound, allowing detailed information regarding the interplay between lattice and magnetic degrees of freedom to be obtained [\cite{Luthi,Luthi2,Bhattacharjee,Watanabe1,Watanabe2,Watanabe3,Nii,Watanabe4,Watanabe5,Watanabe6,Watanabe7,Watanabe8,Watanabe9,Kino,Kataoka,Hazama,Watanabe10}]. As examples, elastic anomalies observed in frustrated spinel $A$Cr$_2$O$_4$ and pseudobrookite CoTi$_2$O$_5$ have identified these compounds as spin-JT antiferromagnets [\cite{Watanabe3,Watanabe10}]. In addition to the case of magnets, in FS-nested metals, the electron-phonon coupling can lead to softening of the sound dispersion due to the Kohn anomaly [\cite{Shaw,Lamago,Powell,Farber,Yang,Zhang,SWang2,Stassis,Isida}]. In the present study of CrB$_2$, we identify two distinct types of unusual temperature dependence in the elastic moduli, observed in symmetrically different elastic modes. One can be attributed to magnetostructural fluctuations (spin-JT fluctuations), while the other can be associated with FS nesting.

\section{Experimental}

High-quality single crystals of CrB$_2$ with $T_N\sim$ 88 K were prepared using the optical floating-zone method as described in Ref. [\cite{Bauer}]. Ultrasound velocities were measured using the phase-comparison technique with longitudinal and transverse sound waves at a frequency of 30 MHz, so that ultrasound velocities or elastic moduli could be measured with a precision of approximately parts per million. The ultrasound waves were generated and detected by LiNbO$_3$ transducers with a fundamental frequency of 30 MHz, which were attached to parallel mirror surfaces of the crystal oriented perpendicular to the hexagonal $a$- and $c$-axes. Measurements were taken to determine the symmetrically independent elastic moduli of the hexagonal crystal, specifically the compressive elastic moduli $C_{11}$ and $C_{33}$, and the shear elastic moduli $C_{44}$ and $C_{66}$. In Fig. 1, the propagation vector {\bf k} and polarization vector {\bf u} of the sound waves corresponding to each elastic modulus are indicated along with the Cr sites of CrB$_2$ in the hexagonal crystal unit cell, which form a triangular lattice in the $ab$-plane. As indicated in Fig. 1, the longitudinal sound wave corresponding to the compressive elastic modulus $C_{11}$ propagates along the hexagonal $a$-axis ({\bf k}$\parallel${\bf u}$\parallel$[100]), whereas the longitudinal wave corresponding to $C_{33}$ propagates along the $c$-axis ({\bf k}$\parallel${\bf u}$\parallel$[001]). Likewise, in Fig. 1, the transverse sound wave corresponding to the shear elastic modulus $C_{44}$ propagates and polarizes in the $ac$-plane ({\bf k}$\parallel$[001] and {\bf u}$\parallel$[100]), whereas the transverse wave corresponding to $C_{66}$ propagates and polarizes in the $ab$-plane ({\bf k}$\parallel$[100] and {\bf u}$\parallel$[120]). The sound velocities of CrB$_2$ measured at a temperature of 300 K are listed on the right side of Fig. 1.

\section{Results}

\begin{figure}[t]
\begin{center}
\includegraphics[scale=0.6]{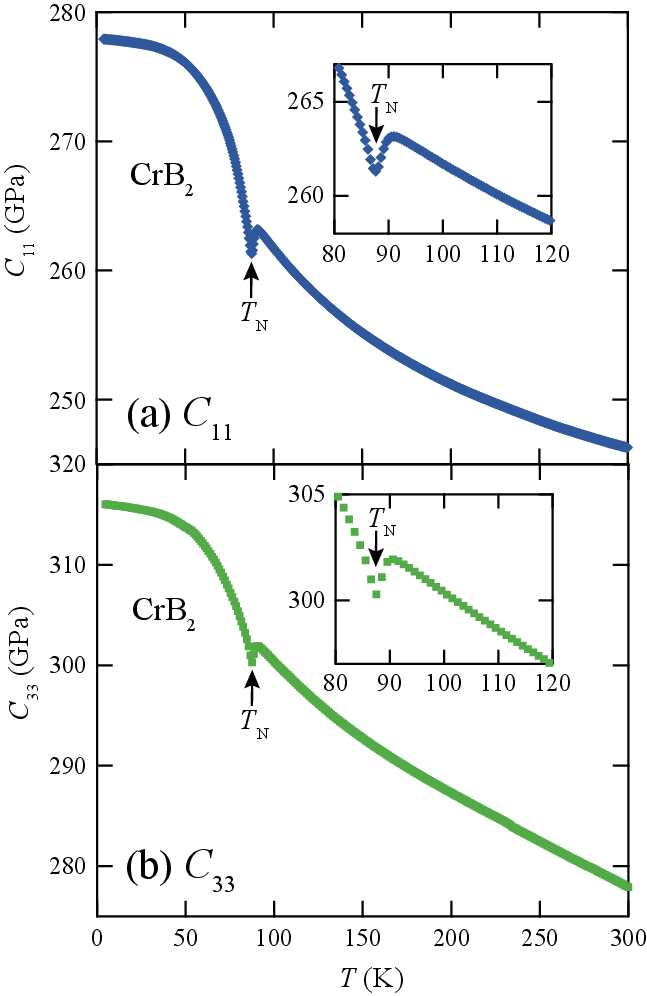}
\caption{\label{fig:fig2} Compressive elastic moduli of CrB$_2$ as functions of $T$: (a) $C_{11}(T)$ and (b) $C_{33}(T)$. The insets in (a) and (b) respectively show expanded views of $C_{11}(T)$ and $C_{33}(T)$ in the range 80 K $\leq T \leq$ 120 K. The labeled arrows indicate $T_N \sim$ 88 K for CrB$_2$.}
\end{center}
\end{figure}

\begin{figure}[t]
\begin{center}
\includegraphics[scale=0.6]{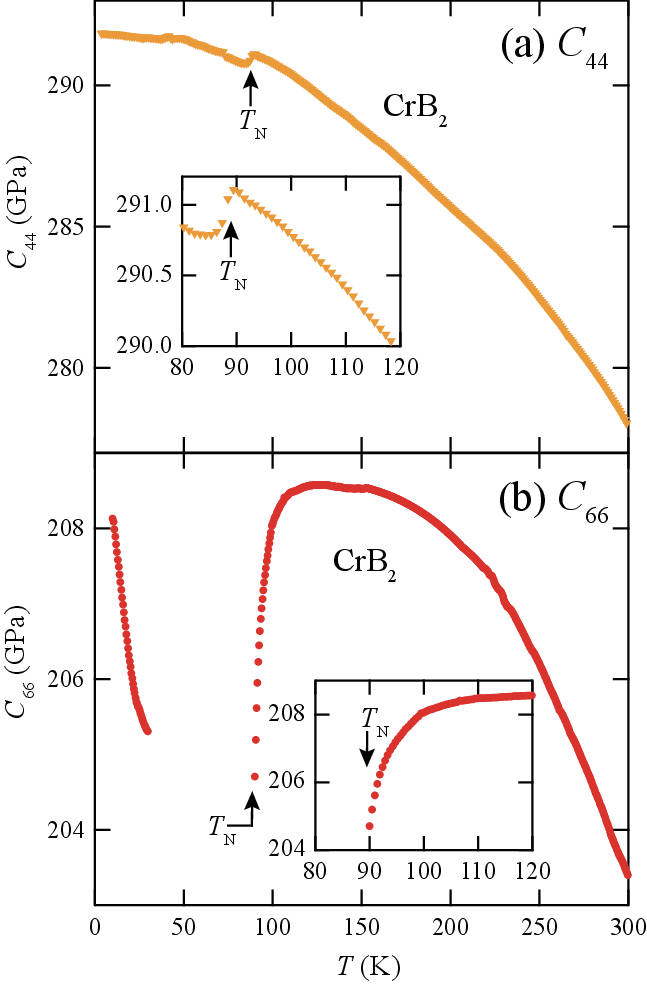}
\caption{\label{fig:fig3} Shear elastic moduli of CrB$_2$ as functions of $T$: (a) $C_{44}(T)$ and (b) $C_{66}(T)$. The insets in (a) and (b) respectively show expanded views of $C_{44}(T)$ and $C_{66}(T)$ in the range 80 K $\leq T \leq$ 120 K. The labeled arrows indicate $T_N \sim$ 88 K for CrB$_2$.}
\end{center}
\end{figure}

Figures 2(a) and 2(b) respectively present the temperature ($T$) dependence of the compressive elastic moduli $C_{11}(T)$ and $C_{33}(T)$ in CrB$_2$. These compressive moduli both harden upon cooling from 300 K to $\sim$90 K with a concave $T$ dependence, exhibit a sharp dip at $T_N\sim$ 88 K, and further harden upon cooling below $T_N$ with a convex $T$ dependence.

Figures 3(a) and 3(b) respectively depict the $T$ dependence of the shear elastic moduli $C_{44}(T)$ and $C_{66}(T)$ in CrB$_2$. These shear moduli both harden upon cooling from 300 K to $\sim$120 K with a convex $T$ dependence, as is usually observed in solids [\cite{Varshni}]. However, below $\sim$120 K, $C_{66}(T)$ [Fig. 3(b)] exhibits Curie-type ($\sim-1/T$-type) softening upon cooling to $T_N\sim$ 88 K, while $C_{44}(T)$ [Fig. 3(a)] hardens monotonically with a convex $T$ dependence. Additionally, $C_{44}(T)$ [Fig. 3(a)] exhibits a small discontinuity at $T_N\sim$ 88 K, and further hardens upon cooling below $T_N$ with a convex $T$ dependence.

For $C_{66}(T)$ [Fig. 3(b)], the ultrasound signal disappears below $T_N$ down to $\sim$40 K, disabling the measurement, and subsequently recovers below $\sim$40 K, signaling elastic hardening of the material upon cooling. The magnitude of the ultrasound signal below $\sim$40 K is smaller than that above $T_N$. In magnetically ordered states, magnetostriction can induce domain-wall stress, resulting in the scattering of ultrasound waves by domain walls [\cite{Bozorth}]. Thus, for $C_{66}(T)$, the ultrasound dissipation below $T_N$ can be attributed to domain-wall-stress effects. The disappearance of the ultrasound signal in the higher-$T$ region of the AF phase ($\sim$40 K $\le T \le T_N$) suggests that thermally activated domain-wall motion enhances ultrasound scattering. In CrB$_2$, the specific heat exhibits an anomaly around $\sim T_N/2$, attributed to unusual excitations other than spin-wave excitations [\cite{Bauer}]. These unusual excitations may be related to the disappearance of the ultrasound signal in the range $\sim$40 K $\le T \le T_N$. In contrast to $C_{66}(T)$, ultrasound dissipation below $T_N$ in $C_{11}(T)$, $C_{33}(T)$, and $C_{44}(T)$ is negligible, indicating that domain-wall-stress effects are very weak for these moduli. This suggests that domain-wall motion couples selectively to the symmetry-lowering $ab$-plane shear strain associated with $C_{66}$.

\section{Discussion}
\subsection{Curie-type softening in shear modulus $C_{66}(T)$}

\begin{figure}[t]
\begin{center}
\includegraphics[scale=0.6]{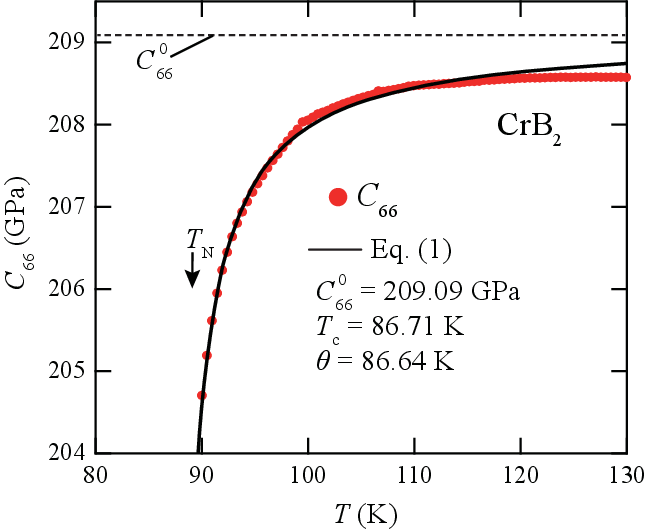}
\caption{\label{fig:fig6} $C_{66}(T)$ of CrB$_2$ in the range 80 K $\leq T \leq$ 130 K [markers, from Fig. 3(b)]. The labeled arrows indicate $T_N \sim$ 88 K for CrB$_2$. The solid curve is a fit of Eq. (1) to the experimental $C_{66}(T)$ in the range $T_N \leq T \leq$ 120 K; the values for the fit parameters are also listed. The dashed horizontal line indicates $C_{66}^{0}$ in Eq. (1).}
\end{center}
\end{figure}

For CrB$_2$, ultrasound velocity measurements revealed Curie-type softening in $C_{66}(T)$ above $T_N$, which ceases at $T_N$ concomitantly with the loss of the ultrasound signal due to the domain-wall-stress effects [Fig. 3(b)]. In solids, Curie-type softening in the $T$ dependence of the elastic modulus $C_{\Gamma}(T)$ emerges as a precursor to a structural transition driven by coupling between the lattice and the electronic degrees of freedom [\cite{Luthi,Luthi2,Kino,Kataoka,Hazama,Nii,Watanabe1,Watanabe2,Watanabe3,Watanabe4,Watanabe7,Watanabe10,Bhattacharjee}]. For CrB$_2$, the observation of a single discontinuous elastic anomaly at $T_N$ in $C_{11}(T)$ [Fig. 2(a)], $C_{33}(T)$ [Fig. 2(b)], and $C_{44}(T)$ [Fig. 3(a)] indicates that only a single phase transition occurs at $T_N$. This excludes the possibility of a structural phase transition occurring at a temperature different from $T_N$. Thus, the Curie-type softening in CrB$_2$ is likely of magnetic origin rather than arising from "nonmagnetic" mechanisms such as orbital Jahn--Teller fluctuations [\cite{Luthi2,Kino,Kataoka,Hazama,Nii,Watanabe4,Watanabe7}]. Therefore, the Curie-type softening observed in the $ab$-plane shear mode $C_{66}(T)$ of CrB$_2$ [Fig. 1] above $T_N$ should be regarded as a precursor to a magnetostructural transition at $T_N$, in which AF ordering coincides with an $ab$-plane shear lattice distortion. In CrB$_2$, such a lattice distortion is consistent with the experimentally observed AF order with {\bf q} = $(0.285, 0.285, 0)$ [\cite{Funahashi,Kaya}], which is expected to arise from lowering the symmetry of the $ab$-plane Cr triangular lattice [Fig. 1]. Consequently, in the frustrated CrB$_2$, the symmetry-lowering magnetostructural transition at $T_N$ should relieve the frustration, corresponding to a spin-JT transition. Thus, the Curie-type softening in $C_{66}(T)$ above $T_N$ is driven by spin-JT fluctuations.

As mentioned in the last paragraph of Sec. III, in CrB$_2$, only the $ab$-plane shear mode $C_{66}(T)$ exhibits pronounced domain-wall-stress effects in the AF phase, whereas such effects are negligible in $C_{11}(T)$, $C_{33}(T)$, and $C_{44}(T)$. This suggests that the magnetostrictive strain associated with magnetic domains corresponds to a volume-preserving $ab$-plane shear distortion that breaks the six-fold symmetry of the Cr triangular lattice, consistent with the spin-JT scenario for CrB$_2$.

In spin-JT systems, the $T$ dependence of the elastic modulus $C_{\Gamma}(T)$ is explained by assuming a coupling between ultrasound and the magnetic atoms through magnetoelastic interactions acting on the exchange interactions, where the exchange striction arises from an ultrasound modulation of the exchange interactions [\cite{Luthi,Watanabe3,Watanabe10}]. In analogy with orbital Jahn--Teller systems [\cite{Luthi,Nii,Watanabe7,Kino,Kataoka,Hazama}], Curie-type softening in $C_{\Gamma}(T)$ in spin-JT systems can be understood by considering the coupling of ultrasound to the magnetic atoms via the exchange-striction mechanism, together with the presence of exchange-striction-sensitive intersite spin--spin interactions [\cite{Watanabe3,Watanabe10}]. The mean-field expression for the Curie-type softening of $C_{\Gamma}(T)$ in spin-JT systems is
\begin{equation}
C_{\Gamma}(T) = C_{\Gamma}^{0} \frac{T-T_c}{T-\theta}.
\label{eq:Curie1}
\end{equation}
Here, $C^{0}_{\Gamma}$ is the background elastic constant, $T_c$ is the second-order critical temperature for which elastic softening $C_{\Gamma}\rightarrow$ 0, and $\theta$ is the exchange-striction-sensitive intersite spin--spin interaction [\cite{Luthi,Watanabe3,Watanabe10}]. The parameter $\theta$ is positive (negative) when the interaction is ferrodistortive (antiferrodistortive).

In Fig. 4, a fit of Eq. (1) to the experimental values of $C_{66}(T)$ for CrB$_2$ in the range $T_N \leq T \leq$ 120 K is shown as a solid curve, with the corresponding fit parameters listed. The fit curve aligns well with the experimental data within this $T$ range. The fit parameter value of $T_c$ is close to the experimentally observed N$\acute{e}$el temperature $T_N$, suggesting that the phase transition at $T_N$ is of second order, which is consistent with specific heat measurements [\cite{Bauer}]. The positive fit value of $\theta$ indicates the dominance of ferrodistortive intersite spin--spin interactions. Furthermore, the magnitude $|\theta|$, which is close to $T_N$, suggests that the magnetostructural transition in CrB$_2$ is governed by intersite spin--spin interactions, consistent with the spin-JT scenario. In Fig. 4, small deviations of the experimental $C_{66}(T)$ data from the fit occur above $\sim$120 K, indicating that the $T$ dependence of the background elastic modulus contributes to the experimental $C_{66}(T)$, which is ignored in Eq. (1) with the assumption of constant $C^0_{\Gamma}$.

For CrB$_2$, the magnitude of the Curie-type softening ($\Delta C_{66}/C_{66}\sim1.9$ $\%$) [Fig. 4] is significantly smaller than that observed in cubic spinel $A$Cr$_2$O$_4$, a prototypical spin-JT system ($\Delta C_{\Gamma}/C_{\Gamma}\sim50$ $\%$) [\cite{Watanabe3}]. However, the magnitude of softening is larger than that in orthorhombic pseudobrookite CoTi$_2$O$_5$, a low-crystal-symmetry spin-JT-system ($\Delta C_{\Gamma}/C_{\Gamma}\sim0.3$ $\%$) [\cite{Watanabe10}]. For CrB$_2$, this significantly smaller softening above $T_N$ compared with that of $A$Cr$_2$O$_4$ is consistent with the fact that the symmetry-lowering lattice distortion below $T_N$ is too small to have been resolved experimentally [\cite{Regnat}], which is similar to the case of CoTi$_2$O$_5$ [\cite{Kirschner,Guratinder}]. The small magnitude of the Curie-type softening might be characteristic of metallic spin-JT systems, in which magnetic frustration arises from competing exchange interactions extending to further-neighbor spins.

\subsection{Softness of compressive moduli $C_{11}(T)$, $C_{33}(T)$}

\begin{figure}[t]
\begin{center}
\includegraphics[scale=0.6]{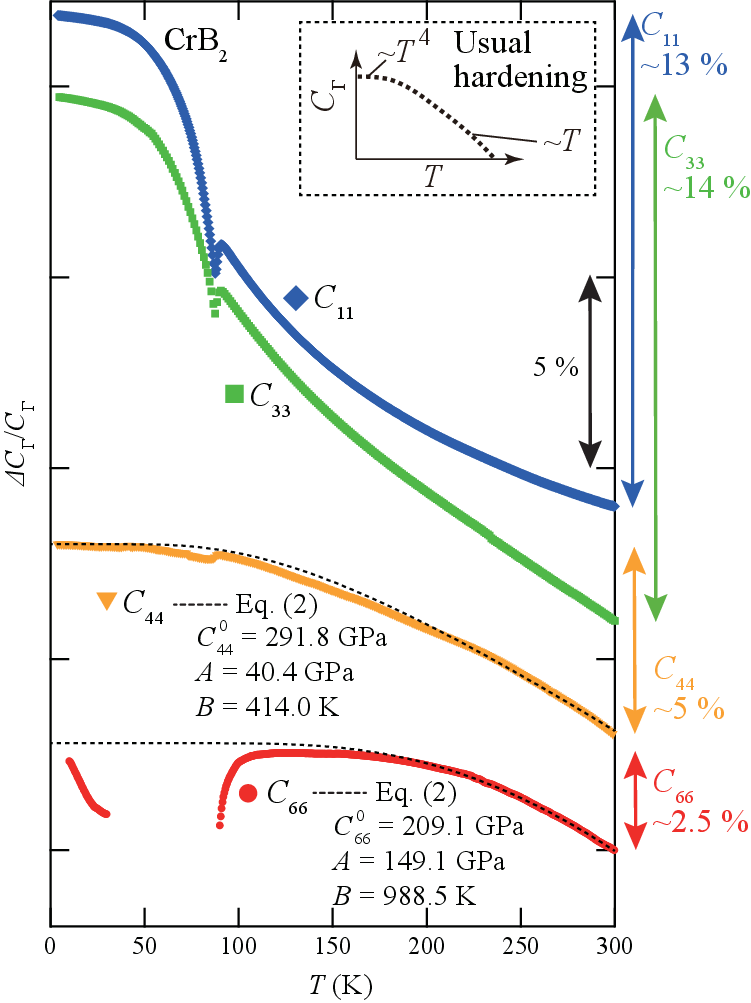}
\caption{\label{fig:fig5} Comparison of the relative shifts of the compressive moduli $C_{11}(T)$ and $C_{33}(T)$, and the shear moduli $C_{44}(T)$ and $C_{66}(T)$ of CrB$_2$ [markers, from Figs. 2 and 3]. The plotted curves have been vertically shifted for clarity. The solid double arrows at the right side of the figure indicate the variation in the experimental $C_{\Gamma}(T)$, which is defined as the difference between the maximum and the minimum values of $C_{\Gamma}(T)$. The minimum occurs at the highest measured $T$= 300 K for all the elastic moduli, and the maximum occurs at the lowest measured $T$ = 2 K for $C_{11}(T)$, $C_{33}(T)$, and $C_{44}(T)$, and at $T\sim$ 120 K for $C_{66}(T)$. The inset is a schematic of $C_{\Gamma}(T)$ typically observed for solids, exhibiting monotonic hardening with decreasing $T$ and a convex $T$ dependence [\cite{Varshni}]. The dashed curves are fits of Eq. (2) to the experimental shear moduli $C_{44}(T)$ and $C_{66}(T)$ in the range 200 K $\leq T \leq$ 300 K; the values for the fit parameters are also listed.}
\end{center}
\end{figure}

For CrB$_2$, while $C_{66}(T)$ [Fig. 3(b)] exhibits Curie-type softening in the range $T_N \leq T \leq \sim$ 120 K, $C_{11}(T)$ [Fig. 2(a)], $C_{33}(T)$ [Fig. 2(b)], and $C_{44}(T)$ [Fig. 3(a)] exhibit hardening upon cooling for $T_N < T \leq$ 300 K. However, the hardening behavior of the compressive moduli $C_{11}(T)$ and $C_{33}(T)$ differs from conventional hardening [\cite{Varshni}]. In typical solids, the elastic modulus hardens linearly with decreasing $T$ at high temperatures, and hardens as $\sim T^4$ at sufficiently low temperatures, which is hardening with a convex $T$ dependence [inset in Fig. 5] [\cite{Varshni}]. In contrast, $C_{11}(T)$ [Fig. 2(a)] and $C_{33}(T)$ [Fig. 2(b)] of CrB$_2$ behave differently in that these moduli exhibit hardening with an unusual concave $T$ dependence for $T_N < T \leq$ 300 K. Additionally, below $T_N$, $C_{11}(T)$ [Fig. 2(a)] and $C_{33}(T)$ [Fig. 2(b)] exhibit another unusual feature: these moduli steeply harden upon cooling below $T_N$ to $\sim$50 K.

Figure 5 compares the relative shifts of the compressive moduli $C_{11}(T)$ and $C_{33}(T)$, and the shear moduli $C_{44}(T)$ and $C_{66}(T)$ of CrB$_2$ [markers, from Figs. 2 and 3]. The solid double arrows at the right side of the figure indicate the magnitude of variation in the experimental $C_{\Gamma}(T)$, which is defined as the difference between the maximum and the minimum in $C_{\Gamma}(T)$. Specifically, the minimum is at the highest measured $T$= 300 K for all the elastic moduli, and the maximum is at the lowest measured $T$ = 2 K in $C_{11}(T)$, $C_{33}(T)$, and $C_{44}(T)$, and at $T\sim$ 120 K in $C_{66}(T)$. In Fig. 5, it is evident that the variations in the compressive moduli $C_{11}(T)$ and $C_{33}(T)$, $\Delta C_{11}/C_{11}\sim$ 13 \% and $\Delta C_{33}/C_{33}\sim$ 14 \%, respectively, are remarkably larger than those in the shear moduli $C_{44}(T)$ and $C_{66}(T)$, $\Delta C_{44}/C_{44}\sim$ 5 \% and $\Delta C_{66}/C_{66}\sim$ 2.5 \%, respectively.

\begin{table}[t]
\caption{\label{tab:table1} Room-temperature experimental values of the elastic moduli $C_{\Gamma}$ (GPa) of AlB$_2$-type hexagonal CrB$_{2}$ [from Figs. 2 and 3, $T$ = 300 K], TiB$_2$ [\cite{Spoor}], and ZrB$_2$ [\cite{Okamoto}]. The ratios $C_{11}/C_{66}$ and $C_{33}/C_{44}$ are also listed.}
\begin{ruledtabular}
\begin{tabular}{ccccccc}
 &$C_{11}$&$C_{33}$&$C_{44}$&$C_{66}$&$C_{11}/C_{66}$&$C_{33}/C_{44}$\\
\hline
CrB$_2$&246&278&278&203&1.21&1.00\\
\hline
TiB$_2$&660&432&260&306&2.16&1.66\\
ZrB$_2$&568&436&248&256&2.22&1.76\\
\end{tabular}
\end{ruledtabular}
\end{table}

For the usual $C_{\Gamma}(T)$ in solids [inset to Fig. 5], an empirical equation known as the Varshni equation provides a good description of the $T$ dependence [\cite{Varshni}]:

\begin{equation}
C_{\Gamma}^0(T)=C_{\Gamma}^0-\frac{A}{\exp(B/T)-1}.
\label{eq:Curie1}
\end{equation}
Here, $C_{\Gamma}^0$ is the elastic modulus at $T$ = 0 K, and $A$ and $B$ are fitting parameters. In Fig. 5, the fits of Eq. (2) to the experimental shear moduli $C_{44}(T)$ and $C_{66}(T)$ in the range 200 K $\leq T \leq$ 300 K are shown as dashed curves. For $C_{44}(T)$, the fit curve reproduces the experimental data in the range 2 K $\leq T \leq$ 300 K, except for the anomaly around $T_N$, and for $C_{66}(T)$, the fit reproduces the experimental data in the range $\sim$120 K $< T \leq$ 300 K, indicating the dominance of the usual $T$ dependence. For the compressive moduli $C_{11}(T)$ and $C_{33}(T)$ [Fig. 5], Eq. (2), which describes the usual behavior of $C_{\Gamma}(T)$ [inset to Fig. 5], fails to reproduce the experimental data.

We note that the experimental absolute values of the room-temperature compressive moduli $C_{11}$ and $C_{33}$ in CrB$_2$ are, respectively, considerably small compared with those of the isostructural hexagonal borides TiB$_2$ and ZrB$_2$, as listed in TABLE I [from Fig. 2] [\cite{Spoor,Okamoto}]. In contrast, the experimental absolute values of the room-temperature shear moduli $C_{44}$ and $C_{66}$ in CrB$_2$, TiB$_2$, and ZrB$_2$ are comparable, as also listed in TABLE I [from Fig. 3] [\cite{Spoor,Okamoto}]. Consequently, the ratios $C_{11}/C_{66}$ and $C_{33}/C_{44}$ in CrB$_2$ are, respectively, considerably smaller than those in TiB$_2$ and ZrB$_2$ [TABLE I].

For $C_{11}(T)$ and $C_{33}(T)$ in CrB$_2$, the large variations [Fig. 5] and the small absolute values at room temperature [TABLE I] indicate that these moduli become unusually softer at higher temperatures, i.e., the longitudinal acoustic phonons become softer at higher temperatures. Considering the absence of this elastic softness in the insulating spin-JT systems of $A$Cr$_2$O$_4$ and CoTi$_2$O$_5$ [\cite{Watanabe3,Watanabe10}], the unusual softness in $C_{11}(T)$ and $C_{33}(T)$ of CrB$_2$ should be related to the metallic nature of this boride.

\begin{figure}[t]
\begin{center}
\includegraphics[scale=0.6]{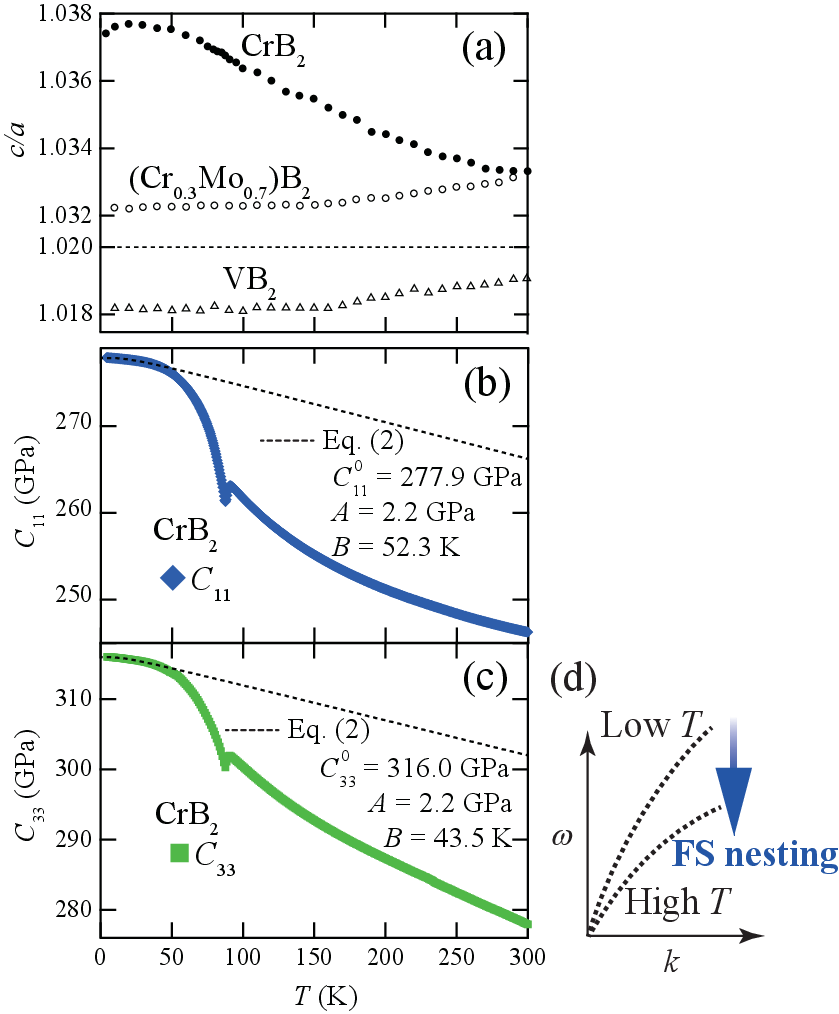}
\caption{\label{fig:fig6} (a) Ratios of hexagonal lattice constants, $c/a$, for the antiferromagnet CrB$_2$ and paramagnets (Cr$_{0.3}$Mo$_{0.7}$)B$_2$ and VB$_2$ as functions of $T$ [from Ref. [\cite{Nishihara}]]. (b) The moduli $C_{11}(T)$ and (c) $C_{33}(T)$ of CrB$_2$ [markers, from Figs. 2(a) and 2(b), respectively]. The dashed curves in (b) and (c) are fits of Eq. (2) to the experimental $C_{\Gamma}(T)$ in the range 2 K $\leq T \leq$ 50 K; the values for the fit parameters are also listed. (d) Schematics of low-$T$ and high-$T$ longitudinal acoustic phonon branches in CrB$_2$, where the high-$T$ branch is softened due to FS nesting as indicated by the arrow.}
\end{center}
\end{figure}

For CrB$_2$, the unusual softness of the compressive moduli $C_{11}(T)$ and $C_{33}(T)$ is naturally explained by FS-nesting-induced phonon softening (Kohn anomaly). Band calculations in CrB$_2$ suggested the presence of FS nesting and predicted the emergence of SDW order along the $c$-axis [\cite{Liu}], whereas SDW-type order was experimentally ruled out [\cite{Funahashi,Kaya}]. In the present study for CrB$_2$, the softness in $C_{11}(T)$ and $C_{33}(T)$ is notably suppressed at lower temperatures, which is compatible with the absence of SDW-type order in CrB$_2$. Therefore, in $C_{11}(T)$ and $C_{33}(T)$ for CrB$_2$, the trend of hardening upon cooling with large variations [Fig. 5] is consistent with the suppression of the FS-nesting instability with decreasing $T$. In this picture, the FS-nesting instability affects the longitudinal acoustic phonons along not only the $a$-axis but also the $c$-axis. In contrast, the behavior of the shear moduli $C_{44}(T)$ and $C_{66}(T)$ [Fig. 5] indicates that the FS-nesting instability is absent or significantly weaker in the transverse acoustic phonons.

We note that, in the antiferromagnet CrB$_2$, the ratio of the hexagonal lattice constants, $c/a$, exhibits stronger $T$ dependence than in the isostructural paramagnets (Cr$_{1-x}$Mo$_x$)B$_2$ and VB$_2$, as reported by thermal expansion measurements [Fig. 6(a)] [\cite{Nishihara}]. The $c/a$ ratios of (Cr$_{1-x}$Mo$_x$)B$_2$ and VB$_2$ decrease only weakly upon cooling, whereas $c/a$ for CrB$_2$ increases upon cooling from 300 K down to $\sim$50 K and then levels off below $\sim$50 K. In CrB$_2$, this increase in $c/a$ originates from the expansion of the $c$-axis and the contraction of the $a$-axis upon cooling [\cite{Nishihara}]. Such $T$-dependent variations of the lattice parameters are expected to modify the FS topology, giving rise to $T$-dependent FS-nesting in CrB$_2$. Furthermore, the increase in $c/a$ upon cooling in the antiferromagnet CrB$_2$, which is not present in the paramagnets (Cr$_{1-x}$Mo$_x$)B$_2$ and VB$_2$, suggests a close relationship with AF ordering.

Figure 6 compares the $T$ dependence of the ratio $c/a$ in CrB$_2$ [Fig. 6(a)] [\cite{Nishihara}] with the compressive moduli $C_{11}(T)$ [Fig. 6(b)] and $C_{33}(T)$ [Fig. 6(c)] for CrB$_2$. The $T$ range exhibiting the strong $T$ dependence for the ratio $c/a$ [Fig. 6(a)], $\sim$50 K $< T <$ 300 K, is in agreement with those for $C_{11}(T)$ [Fig. 6(b)] and $C_{33}(T)$ [Fig. 6(c)], for which FS-nesting is expected to lead to $T$ dependence. The weak $T$ dependence of the ratio $c/a$, $C_{11}(T)$, and $C_{33}(T)$ below $\sim$50 K suggests that $T$-dependent FS-nesting is absent or weak below $\sim$50 K. However, it is notable that, for $C_{11}(T)$ [Fig. 6(b)] and $C_{33}(T)$ [Fig. 6(c)], the experimental values at the lowest $T$ = 2 K of $C_{11}\sim$ 278 GPa and $C_{33}\sim$ 316 GPa are, respectively, still considerably smaller than the room-temperature values of $C_{11}$ and $C_{33}$ in TiB$_2$ and ZrB$_2$ [TABLE I], which suggests that FS-nesting is present even below $\sim$50 K, although its $T$ dependence is absent or weak.

For CrB$_2$, since $T$-dependent FS-nesting should be absent or weak below $\sim$50 K, $C_{11}(T)$ and $C_{33}(T)$ below $\sim$50 K are expected to be well modeled by the usual $T$ dependence described in Eq. (2) [\cite{Varshni}]. In Figs. 6(b) and 6(c), the fits of Eq. (2) to the experimental $C_{\Gamma}(T)$ for $T \leq$ 50 K are shown as dashed curves. The fit curves deviate from the experimental $C_{\Gamma}(T)$ above $\sim$50 K, which becomes larger at higher $T$. This deviation suggests that the longitudinal acoustic phonons become softer at higher $T$ because of FS nesting, as illustrated in Fig. 6(d).

\subsection{Magnetoelastic effects in CrB$_2$}

\begin{figure}[t]
\begin{center}
\includegraphics[scale=0.6]{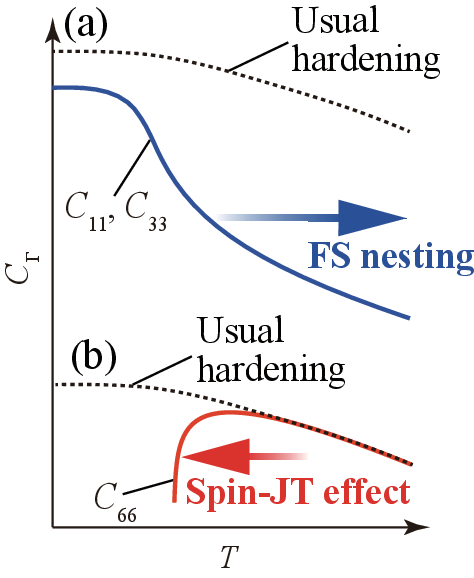}
\caption{\label{fig:fig7} Schematics of (a) compressive moduli $C_{11}(T)$ and $C_{33}(T)$ of CrB$_2$, and (b) shear modulus $C_{66}(T)$ of CrB$_2$. The dashed curves represent the $C_{\Gamma}(T)$ typically observed for solids, exhibiting monotonic hardening with decreasing $T$ and a convex $T$ dependence [\cite{Varshni}]. The arrow in (a) indicates softening in $C_{11}(T)$ and $C_{33}(T)$ of CrB$_2$ at higher temperatures driven by FS nesting. The arrow in (b) indicates softening in $C_{66}(T)$ of CrB$_2$ at lower temperatures, i.e., Curie-type softening due to spin-JT effect.}
\end{center}
\end{figure}

Figure 7 schematically summarizes the unusual $C_{\Gamma}(T)$ of CrB$_2$ discussed in Secs. IV A and IV B. In Fig. 7(a), the softness in the compressive moduli $C_{11}(T)$ and $C_{33}(T)$ is attributed to the $T$-dependent FS-nesting instability [Sec. IV B], and in Fig. 7(b), the Curie-type softening in the shear modulus $C_{66}(T)$ above $T_N$, serving as a precursor to the spin-JT transition at $T_N$ [Sec. IV A], is illustrated.

As discussed in Sec. IV B, the behavior of $C_{11}(T)$ and $C_{33}(T)$ in CrB$_2$ [Fig. 7(a)] is consistent with the presence of an FS-nesting instability in the longitudinal acoustic phonons, together with its suppression upon cooling concomitant with an increase in the lattice-constant ratio $c/a$ [Fig. 6(a)]. In the antiferromagnet CrB$_2$, this increase in $c/a$ suggests a close relationship with AF ordering, whereas no such increase is observed in the paramagnets (Cr$_{1-x}$Mo$_x$)B$_2$ and VB$_2$ [Fig. 6(a)]. Notably, while hydrostatic pressure suppresses $T_N$ in CrB$_2$ [\cite{Pei}], uniaxial pressure experiments have shown that $T_N$ increases with increasing $c/a$ [\cite{Regnat}]. Therefore, the increase in $c/a$ upon cooling in CrB$_2$ [Fig. 6(a)] is likely driven by longitudinal magnetoelastic coupling, which enhances the exchange interactions relevant to AF ordering while simultaneously suppressing the FS-nesting instability in longitudinal acoustic phonons.

For CrB$_2$, band structure calculations combined with de Haas--van Alphen experiments revealed that the Cr-3$d$-derived FS sheets are highly sensitive to magnetic order, whereas the B-2$p$-derived sheets are insensitive [\cite{Brasse}]. This suggests that the Cr-3$d$-derived FS sheets are likely responsible for the FS nesting relevant to AF ordering in CrB$_2$. In addition, band structure calculations showed that the closed FS sheets partly expand with decreasing $c/a$ [\cite{Brasse}], consistent with the $c/a$-dependent FS nesting suggested in Sec. IV B and Fig. 6. Thus, the ratio $c/a$ may serve as a tuning parameter for the magnetic and electronic states in CrB$_2$, potentially relevant to the pressure-induced superconductivity proposed in this system [\cite{Biswas,Pei}].

In CrB$_2$, as discussed above, the exchange interactions are expected to be enhanced upon cooling concomitantly with the increase in $c/a$ [Fig. 6(a)] and consequently become frustrated on the Cr triangular lattice. As discussed in Sec. IV A, the Curie-type softening observed in the shear modulus $C_{66}(T)$ above $T_N$ [Fig. 7(b)] suggests that this frustration is relieved via a spin-JT transition, in which the symmetry of the Cr triangular lattice is lowered through $ab$-plane transverse magnetoelastic coupling. Since the $ab$-plane shear modulus $C_{66}$ in CrB$_2$ corresponds to a rotational-symmetry-breaking elastic mode, the observed magnetoelastic fluctuations may reflect nematic-like fluctuations associated with bond anisotropy arising from localized-spin correlations rather than purely itinerant electrons. Further studies are desirable to clarify the possible relationship between the spin-JT instability and spin-driven nematicity in CrB$_2$.

Overall, the present results indicate that longitudinal and transverse magnetoelastic couplings play distinct and significant roles in the metallic frustrated magnetism of CrB$_2$.

\section{Summary}

Ultrasound velocity measurements of the hexagonal metallic frustrated antiferromagnet CrB$_2$ reveal two distinct types of anomalous elastic softness: one in the $ab$-plane shear modulus $C_{66}(T)$ above $T_N$, and the other in the compressive moduli $C_{11}(T)$ along the $a$-axis and $C_{33}(T)$ along the $c$-axis. The Curie-type softening observed in $C_{66}(T)$ above $T_N$ indicates the presence of magnetostructural fluctuations (spin-JT fluctuations), serving as a precursor to a symmetry-lowering lattice distortion at $T_N$ that relieves magnetic frustration. The softness in $C_{11}(T)$ and $C_{33}(T)$ is naturally explained by FS nesting, which is suppressed upon cooling. The present results suggest that, in CrB$_2$, longitudinal magnetoelastic coupling suppresses FS nesting and enhances frustrated exchange interactions, while transverse magnetoelastic coupling plays a key role in relieving the frustration. Details of not only acoustic but also optical phonon behaviors are expected to be studied to further understand the magnetoelastic effects of CrB$_2$.

\section{Acknowledgments}

This study was partly supported by a Grant-in-Aid for Scientific Research (C) (Grant No. 21K03476) from MEXT of Japan. This study received funding from the Deutsche Forschungsgemeinschaft (DFG, German Research Foundation) under TRR80 (From Electronic Correlations to Functionality, Project No. 107745057), TRR360 (Constrained Quantum Matter, Project No. 492547816, Projects C1 and C3), SPP2137 (Skyrmionics, Project No. 403191981, Grant PF393/19), and the excellence cluster MCQST under Germany's Excellence Strategy EXC-2111 (Project No. 390814868). Financial support by the European Research Council (ERC) through Advanced Grants No. 291079 (TOPFIT) and No. 788031 (ExQuiSid) is gratefully acknowledged.

\end{document}